\documentclass[amssymb,prd,aps,amsmath,nofootinbib,twocolumn,preprintnumbers,showpacs,showkeys]{revtex4}

\usepackage{color}
\usepackage{graphicx}
\usepackage{hyperref}
\hypersetup{colorlinks,bookmarksopen,bookmarksnumbered,citecolor=blue,
linkcolor=black,pdfstartview=FitH,urlcolor=blue}
\usepackage{multirow}
\usepackage[utf8]{inputenc}
\usepackage{url}

\usepackage{bm}	
\usepackage{slashed}	
\usepackage{}	
\usepackage{}	

\begin{document}

\title{Minimal dark matter in $SU(5)$ grand unification}

\preprint{KANAZAWA-24-09}

\author{Takashi Toma}

\email{toma@staff.kanazawa-u.ac.jp}

\affiliation{
Institute of Liberal Arts and Science, Kanazawa University, Kanazawa 920-1192, Japan
} 
\affiliation{
Institute for Theoretical Physics, Kanazawa University, Kanazawa 920-1192, Japan
}

\begin{abstract}
Minimal dark matter is an attractive candidate for dark matter because it is stabilized without the need to impose additional symmetries. 
It is known that the $SU(2)_L$ quintuplet fermion can serve as a minimal dark matter candidate, with its mass predicted to be around $14~\mathrm{TeV}$, based on the thermal production mechanism.
In this work, we embed the quintuplet dark matter within nonsupersymmetric $SU(5)$ grand unified theories. 
We find that two pairs of colored sextet fermions are required at the $\mathcal{O}(1-10)~\mathrm{TeV}$ scale to achieve gauge coupling unification,
 with the unification scale near the reduced Planck scale. 
These colored sextet fermions become metastable because their interactions are suppressed by the unification scale. 
Our model can be tested through comprehensive searches for colored sextet fermions in collider experiments, as well as through indirect and direct detection methods for minimal dark matter.
Once the minimal dark matter scenario has been experimentally confirmed, it will have implications for modifying string theories. 
\end{abstract}

\date{\today}


\maketitle

\section{Introduction}
Dark matter is known to exist in the Universe, but its fundamental properties, such as mass, interactions, and production mechanism, remain unknown.
One of the most promising candidates for dark matter is thermal dark matter, which is produced through the freeze-out mechanism.
Despite extensive searches via direct detection, indirect detection, and collider experiments, no signals have been detected so far.
Consequently, experimental constraints on interactions with Standard Model (SM) particles have become increasingly stringent.
In particular, direct detection experiments place severe limits on the spin-independent cross section, with the current strongest bound provided by the LUX-ZEPLIN (LZ) experiment~\cite{LZ:2022lsv, tevpa2024lz}.
Even under these constraints, thermal dark matter scenarios remain appealing due to their strong predictiveness and the successful description of the thermal history of the universe below MeV scales.
For example, candidates such as minimal dark matter~\cite{Cirelli:2005uq}, including the wino in supersymmetric models in a broader sense, 
pseudo-Nambu-Goldstone dark matter~\cite{Gross:2017dan}, and pseudoscalar interacting fermionic dark matter~\cite{Freytsis:2010ne, Abe:2018emu}, 
continue to align with current experimental and observational bounds.

Minimal dark matter, in particular, can be naturally incorporated through simple extensions of the SM, 
such as introducing an $SU(2)_L$ quintuplet fermion or a septuplet scalar~\cite{Cirelli:2005uq}.
The interactions of these particles with the SM are highly constrained by gauge symmetries, 
leading to an accidental $\mathbb{Z}_2$ symmetry that stabilizes the lightest neutral state in the quintuplet or septuplet at the renormalizable level.

Separately, models that unify the SM field content and gauge interactions known as grand unified theories (GUTs) have been studied for decades.
The $SU(5)$ symmetry is well established as the minimal GUT framework, embedding all SM matter particles within the $\overline{\bm{5}}$ and $\bm{10}$ representations~\cite{Georgi:1974sy}.
However, it is also well known that gauge coupling unification is not realized in the minimal $SU(5)$ GUT model when considering only SM particles.

In this work, we propose a simple embedding of the quintuplet minimal dark matter and all other matter particles 
into the $SU(5)$ gauge symmetry.\footnote{Alternative attempts to embed minimal dark matter in $SO(10)$ GUTs have been explored~\cite{Cho:2021yue}, 
and a comprehensive study of dark matter stability in $SO(10)$ unification has also been conducted~\cite{Ferrari:2018rey}.}
We assume that the scale of new light particles in the multiplets is fixed by the dark matter mass, 
approximately $14~\mathrm{TeV}$, to reproduce the thermal relic abundance via the freeze-out mechanism, 
including the effects of Sommerfeld enhancement~\cite{Hisano:2003ec, Hisano:2004ds, Hisano:2006nn} and bound state formation~\cite{Mitridate:2017izz}.
To achieve gauge coupling unification, we will find that two pairs of additional $SU(3)_C$ sextet and antisextet fermions, which can be embedded in $\bm{15}$ and $\overline{\bm{15}}$ representations, are required.
These exotic sextet fermions are metastable due to interactions suppressed by the unification scale and can form bound states with quarks and gluons, which may be detectable in collider experiments.
Together with gamma-ray observations and direct detection experiments for the quintuplet fermion dark matter, these collider searches provide a comprehensive test of our unified model.
In addition, we will discuss how to achieve the large mass splittings between the minimal dark matter mass, which is around the $14~\mathrm{TeV}$ scale, and the unification scale.

\section{Minimal dark matter in $SU(5)$}

The Lagrangian for the fermionic $SU(2)_L$ quintuplet $\chi$ is given by
\begin{align}
\mathcal{L} = \frac{1}{2} \overline{\chi} \left( i \gamma^{\mu}D_{\mu} - M_5 \right) \chi,
\end{align}
where $M_5$ is the tree-level mass~\cite{Cirelli:2005uq}. 
The other interactions are forbidden at the renormalizable level by the SM gauge symmetry. 
An accidental $\mathbb{Z}_2$ symmetry ensures the stability of the lightest component in the quintuplet.
At tree level, all components in the quintuplet are degenerate in mass. 
However, small mass splitting between the charged and neutral components ($\approx166~\mathrm{MeV}$) arises at the loop level~\cite{Cirelli:2005uq}. 
Consequently, the neutral component $\chi^0$ becomes the lightest state and is identified as the dark matter.

The mass $M_5$ is the sole parameter of minimal dark matter, as all interactions are determined by the gauge symmetry. 
The dark matter mass is constrained by the requirement to reproduce the observed relic abundance through the thermal freeze-out mechanism. 
The dominant annihilation channels are $\chi^0\chi^0 \to W^+W^-,ZZ,Z\gamma$ and $\gamma\gamma$, mediated by gauge bosons. 
Accounting for the effects of Sommerfeld enhancement and bound state formation, the mass is determined to be $M_5=14~\mathrm{TeV}$~\cite{Mitridate:2017izz}.

To further investigate, we propose embedding the minimal dark matter within the $SU(5)$ gauge symmetry.
In the Georgi-Glashow model~\cite{Georgi:1974sy}, all SM fermions in a single generation are embedded in the $\overline{\bm{5}}_F$ and $\bm{10}_F$ representations, 
while the SM Higgs doublet $H$ resides in the $\bm{5}_H$ representation.
Additionally, the $\bm{24}_H$ adjoint Higgs field is introduced to achieve the symmetry breaking pattern $SU(5) \to SU(3)_C \times SU(2)_L \times U(1)_Y\equiv G_\mathrm{SM}$ 
via its vacuum expectation value.
Beyond the Georgi-Glashow model, we introduce a $\bm{200}_F$ fermion representation, which is the smallest representation containing an $SU(2)_L$ quintuplet. 
The $\bm{200}_F$ representation can be decomposed under the $SU(3)_C \times SU(2)_L \times U(1)_Y$ subgroup as follows~\cite{Yamatsu:2015npn, Feger:2019tvk, Fonseca:2020vke}: 
\begin{align}
\bm{200}_F=&\hspace{0.05cm}
(\bm{6},\bm{3})_{-5/3} + 
(\bm{15},\bm{2})_{-5/6} + 
(\bm{3},\bm{4})_{-5/6} + 
(\bm{3},\bm{2})_{-5/6}\nonumber\\
&+(\bm{27},\bm{1})_{0} + 
(\bm{8},\bm{3})_{0} + 
(\bm{8},\bm{1})_{0}\nonumber\\
&+(\bm{1},\bm{5})_{0} + 
(\bm{1},\bm{3})_{0} + 
(\bm{1},\bm{1})_{0}\nonumber\\ 
&
+(\overline{\bm{6}},\bm{3})_{5/3}
+(\overline{\bm{15}},\bm{2})_{5/6}
+(\overline{\bm{3}},\bm{4})_{5/6}
+(\overline{\bm{3}},\bm{2})_{5/6},
\label{eq:200}
\end{align}
where we note that the $SU(3)_C$ sextet $\bm{6}$ is defined by the Dynkin label $(0,2)$, following a convention commonly used in particle physics. 
The introduction of the $\bm{200}_F$ representation ensures the incorporation of the quintuplet minimal dark matter into the $SU(5)$ framework, 
providing a natural extension of the Georgi-Glashow model. Further exploration of this embedding will allow for a comprehensive analysis of its phenomenological implications, 
including gauge coupling unification and the search for additional states predicted by the model.

\section{Gauge coupling unification}

To investigate gauge coupling unification, we compute the $\beta$ functions for the gauge couplings $g_i~(i=1,2,3)$ at the two-loop level where 
the gauge coupling for $U(1)_Y$ is normalized as $g_1=\sqrt{5/3}g_Y$. 
The renormalization group equation (RGE) for the gauge couplings in the SM is expressed as~\cite{Machacek:1983tz, Machacek:1983fi, Machacek:1984zw, Arason:1991ic, Luo:2002ey}
\begin{align} 
\frac{dg_i}{dt} = \frac{b_i g_i^3}{(4\pi)^2} +\sum_{j=1}^{3}\frac{b_{ij} g_i^3 g_j^2}{(4\pi)^4} +\frac{c_i g_i^3 y_t^2}{(4\pi)^4}, 
\end{align} 
where $t=\log(\mu/m_Z)$, $y_t$ is the top Yukawa coupling, and the coefficients $b_i$, $b_{ij}$, and $c_i$ are given by
\begin{align}
b_{i}=&\left(
\begin{array}{c}
 41/10\\
-19/6\\
-7
\end{array}
\right),\hspace{0.2cm}
b_{ij}=\left(
\begin{array}{ccc}
199/50 & 27/10 & 44/5 \\
9/10 & 35/6 & 12\\
11/10 & 9/2 & -26
\end{array}
\right),\\
c_i=&\left(
\begin{array}{c}
 -17/10\\
-3/2\\
-2
\end{array}
\right).
\end{align}
The contributions of other Yukawa couplings are neglected due to their small magnitudes compared to $y_t$.

The RGE for the top Yukawa coupling $y_t$ is given by 
\begin{align}
\frac{dy_t}{dt}=&~
\frac{b_{y_t}^{(1)}}{(4\pi)^2}+\frac{b_{y_t}^{(2)}}{(4\pi)^4},
\end{align}
where the one-loop $b_{y_t}^{(1)}$ and two-loop $b_{y_t}^{(2)}$ contributions are
\begin{align}
b_{y_t}^{(1)}=&~
\frac{9}{2}y_t^3-y_t\left(\frac{17}{20}g_1^2+\frac{9}{4}g_2^2+8g_3^2\right),\\
b_{y_t}^{(2)}=&~
y_t\left(-12y_t^4 - 6y_t^2\lambda + \frac{3}{2}\lambda^2\right)\nonumber\\
&+y_t^3\left(\frac{393}{80}g_1^2+\frac{225}{16}g_2^2+36g_3^2\right)\nonumber\\
&+y_t\left(\frac{1187}{600}g_1^4-\frac{9}{20}g_1^2g_2^2+\frac{19}{15}g_1^2g_3^2\right.\nonumber\\
&\left.\hspace{1cm}-\frac{23}{4}g_2^4+9g_2^2g_3^2-108g_3^4\right).
\end{align}
Here, $\lambda$ is the Higgs self-coupling normalized as $\mathcal{L}\supset -\frac{\lambda}{2}|H|^4$. 

The RGE for the Higgs self-coupling $\lambda$ is given by
\begin{align}
\frac{d\lambda}{dt}=&~
\frac{b_\lambda^{(1)}}{(4\pi)^2}+\frac{b_\lambda^{(2)}}{(4\pi)^4},
\end{align}
where the one-loop $b_{\lambda}^{(1)}$ and two-loop $b_{\lambda}^{(2)}$ contributions are
\begin{align}
b_{\lambda}^{(1)}=&~
12\left(\lambda^2+y_t^2\lambda-y_t^4\right) - \lambda\left(\frac{9}{5}g_1^2+9g_2^2\right)\nonumber\\
&+\frac{9}{4}\left(\frac{3}{25}g_1^4+\frac{2}{5}g_1^2g_2^2+g_2^4\right),\\
b_{\lambda}^{(2)}=&
-78\lambda^3-72\lambda^2 y_t^2
+18\lambda^2\left(\frac{3}{5}g_1^2+3g_2^2\right)\nonumber\\
&-3\lambda y_t^4
+10\lambda y_t^2\left(\frac{17}{20}g_1^2+\frac{9}{4}g_2^2+8g_3^2\right)
\nonumber\\
&
-\lambda\left(\frac{73}{8}g_2^4-\frac{117}{20}g_1^2g_2^2-\frac{1887}{200}g_1^4\right)\nonumber\\
&-\frac{3411}{1000}g_1^6-\frac{289}{40}g_1^2g_2^4-\frac{1677}{200}g_1^4g_2^2+\frac{305}{8}g_2^6\nonumber\\
&
+60y_t^6
-\frac{16}{5}y_t^4\left(g_1^2+20g_3^2\right)\nonumber\\
&-\frac{9}{50}y_t^2\left(19g_1^4-70g_1^2g_2^2+25g_2^4\right).
\end{align}

The quintuplet introduces additional contributions to the $\beta$ functions in the energy scale above its mass $M_5$. 
These corrections are given by
\begin{align}
\Delta b_2=&~\frac{20}{3},\hspace{0.5cm}
\Delta b_{22}=\frac{560}{3},\\
\Delta b_{y_t}^{(2)}=&~5g_2^4,\quad
\Delta b_{\lambda}^{(2)}=-2g_2^4\left(4g_1^2+20g_2^2-25\lambda\right),
\end{align}
where we used \texttt{SARAH} to calculate the $\beta$ functions at the two loop level~\cite{Staub:2008uz, Staub:2013tta}. 
These contributions indicate that the gauge couplings do not unify at any scale. 
One may consider whether other states in the $\bm{200}_F$ representation in Eq.~(\ref{eq:200}) such as $(\bm{6},\bm{3})_{-5/3}$ and $(\bm{15},\bm{2})_{-5/6}$ 
could modify the gauge coupling running and achieve gauge coupling unification if they are also light.
However, we have checked that this is impossible. 
In fact, no additional $SU(2)_L$ fields can be introduced, even for the lowest dimensional representation. 
Otherwise, a Landau pole for the gauge coupling $g_2$ would appear below the Planck scale.

To achieve gauge coupling unification, two pairs of $SU(3)_C$ sextet fermions are introduced: $(\bm{6},\bm{1})_{-2/3}$ and $(\overline{\bm{6}},\bm{1})_{2/3}$. 
These fermions can originate from the $\bm{15}_F$ and $\overline{\bm{15}}_F$ representations of $SU(5)$, 
whose branching rules under $G_\mathrm{SM}$ are given by~\cite{Yamatsu:2015npn}
\begin{align}
 \bm{15}_F =&~ (\bm{6},\bm{1})_{-2/3} + (\bm{3},\bm{2})_{1/6} + (\bm{1},\bm{3})_{1},\\
 \overline{\bm{15}}_F =&~ (\overline{\bm{6}},\bm{1})_{2/3} + (\overline{\bm{3}},\bm{2})_{-1/6} + (\bm{1},\bm{3})_{-1}.
\end{align}
A pair of sextet fermions contributes to the $\beta$-functions as follows: 
\begin{align}
\Delta b_1=&~\frac{32}{15},\quad
\Delta b_{11}=\frac{128}{75},\quad
\Delta b_{13}=\frac{64}{3},\\
\Delta b_3=&~\frac{10}{3},\quad
\Delta b_{31}=\frac{8}{3},\quad\hspace{0.4cm}
\Delta b_{33}=\frac{250}{3},\\
\Delta b_{y_t}^{(2)}=&~
\frac{232}{225}g_1^4+\frac{200}{9}g_3^4,\\
\Delta b_{\lambda}^{(2)}=&~
-\frac{16}{125}g_1^4\left(12g_1^2+20g_2^2-25\lambda\right).
\end{align}
The masses of these sextet fermions are denoted as $M_6$ and $\tilde{M}_6$. 
The contributions to the $\beta$-functions are included at the energy scales $\mu=M_6$ and $\mu=\tilde{M}_6$, respectively.

We numerically solve the RGEs at the two-loop level. 
The integration is performed over the energy range $m_Z \leq \mu \leq M_U$, where $M_U$ is the unification scale.
The unified gauge coupling $\alpha_U^{-1}$ at the unification scale $M_U$ is determined by minimizing the quantity~\cite{Schwichtenberg:2018cka}
\begin{align}
\Delta \equiv\sqrt{\Delta \alpha_{12}^{-2}(M_U)+\Delta\alpha_{23}^{-2}(M_U)},
\end{align}
where $\Delta \alpha_{ij}^{-1} = \alpha_{i}^{-1}-\alpha_{j}^{-1}$, and $\alpha_i=g_{i}^2/(4\pi)$ are the redefined gauge couplings. 
The gauge couplings at the reference scale $\mu=m_Z$ are determined using physical constants from experimental measurements~\cite{ParticleDataGroup:2022pth}. 
Specifically: $\alpha_1^{-1}=58.81$, $\alpha_2^{-1}=29.48$, and $\alpha_3^{-1}=8.475$. 
These values are derived using the following physical constants: 
\begin{align}
m_W=&~80.377~\mathrm{GeV},\quad
G_F=1.1663788~\mathrm{GeV}^{-2},\nonumber\\
\sin^2\theta_W=&~0.23121,\hspace{1.3cm}
\alpha_s=0.1180.\nonumber
\end{align}

\begin{figure*}
\begin{center}
 \includegraphics[width=8.5cm]{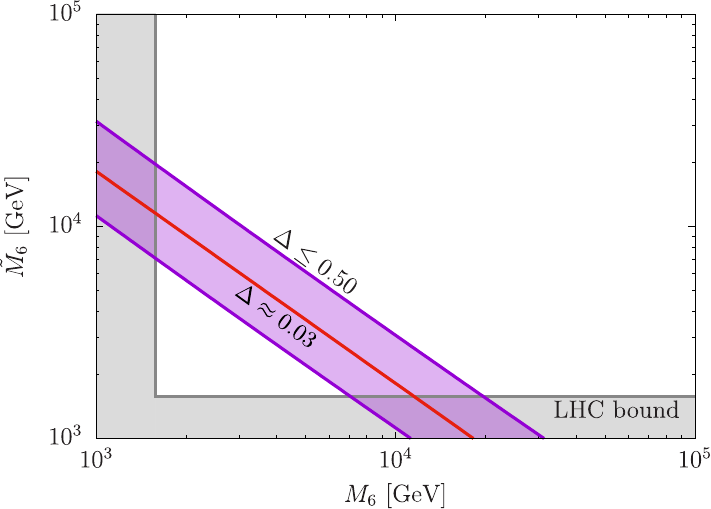}
 \includegraphics[width=8.5cm]{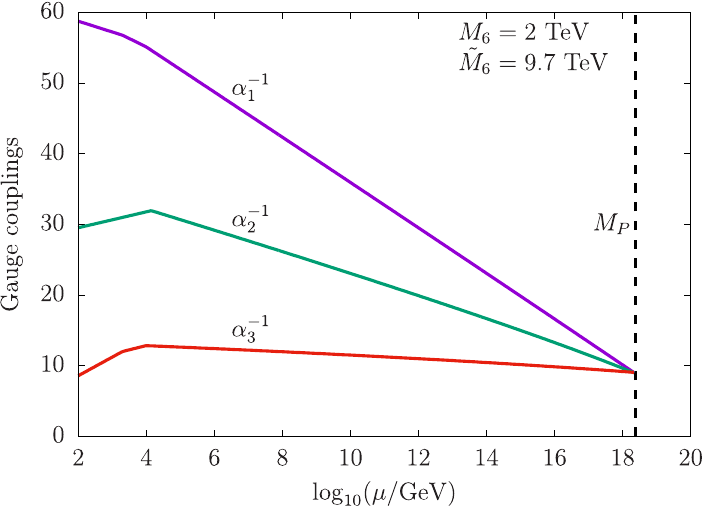}
\caption{Left: contours of $\Delta$ for gauge coupling unification in $M_6$-$\tilde{M}_6$ plane. The red line indicates the region where $\Delta$ is minimized, representing the best gauge coupling unification, 
and the purple region represents the acceptable unification. 
The gray region is excluded based on current experimental bounds from colored fermion searches at the LHC~\cite{Carpenter:2021rkl}. 
Right: example of the gauge coupling running where the sextet fermion masses are fixed to be $M_6=2~\mathrm{TeV}$ and $\tilde{M}_6=9.7~\mathrm{TeV}$. 
The gauge couplings unify at a scale $M_U\approx2.1 \times 10^{18}~\mathrm{GeV}$, close to the Planck scale, and the unified gauge coupling is predicted as $\alpha_U^{-1}\approx9.0$.}
\label{fig:beta}
\end{center}
\end{figure*}

The contours of $\Delta$ are shown in the left panel of Fig.~\ref{fig:beta} where the quintuplet fermion mass is fixed to be $M_5=14~\mathrm{TeV}$. 
The optimal unification ($\Delta\approx0.03$) is achieved when the sextet fermion masses satisfy the relation
\begin{align}
\sqrt{M_6\tilde{M}_6}\approx4.3~\mathrm{TeV}.
\end{align}
Under this condition, the unification scale and the unified gauge coupling are determined to be $M_U=2.1\times10^{18}~\mathrm{GeV}$ and 
$\alpha_U^{-1}=9.0$.\footnote{In fact, there is very small preference for the unification at $M_6=\tilde{M}_6=4.3~\mathrm{TeV}$. However the difference is invisible.} 
Notably, the unification scale is very close to the reduced Planck scale $M_P=2.4\times10^{18}~\mathrm{GeV}$.\footnote{A possibility of 
gauge coupling unification at the Planck scale has been studied~\cite{Howl:2007hq}. 
Additionally, a mechanism for neutrino mass generation from the Planck scale has also been explored~\cite{Ibarra:2018dib, Ibarra:2020eia}.} 
At such high energy scales, quantum gravity effects are expected to become significant and could influence the coupling unification. 
Indeed, the energy scale $E_\mathrm{new}$ at which tree-level unitarity breaks down can be estimated as $E_\mathrm{new}^2=20(G_NN)^{-1}$ where $G_N$ is the Newton constant and 
$N=\frac{2}{3}N_s+N_f+4N_V$ accounts for the number of complex scalars, Weyl fermions and real vector bosons~\cite{Han:2004wt}. 
In our model, we can obtain $E_\mathrm{new}\approx2.7\times10^{18}~\mathrm{GeV}$, which is slightly exceeding the unification scale. 
In the purple region ($\Delta\leq 0.50$), the gauge coupling unification could still be acceptable. 
We have checked that the unification scale and the unified gauge coupling are almost constant along with the red line, and 
they change within the range of $1.7\times10^{18}~\mathrm{GeV}\lesssim M_U \lesssim 2.8\times10^{18}~\mathrm{GeV}$ and $8.4\lesssim \alpha_U^{-1}\lesssim 9.6$ in the purple region. 
An example running of the gauge couplings is shown in the right panel of Fig.~\ref{fig:beta} where the sextet fermion masses are fixed to be $M_6=2~\mathrm{TeV}$ and $\tilde{M}_6=9.7~\mathrm{TeV}$, respectively.
It should be noted that threshold corrections from heavy particles could influence the running of the gauge couplings. However, incorporating these effects is beyond the scope of this work.

Using the unification scale $M_U$ and unified gauge coupling strength $\alpha_U^{-1}$, 
the proton lifetime can be roughly estimated as $\tau_p=\alpha_U^{-2}M_U^4/m_p^5\approx3.9\times10^{39}~\mathrm{years}$ where $m_p$ is the proton mass. 
The predicted lifetime is significantly longer than the current experimental upper bound $\tau_p^\mathrm{exp}=2.4\times10^{34}~\mathrm{years}$ for the $p\to e^+\pi^0$ decay channel, as reported by 
Super-Kamiokande~\cite{Super-Kamiokande:2020wjk}. 
Furthermore, it exceeds the expected future sensitivity of Hyper-Kamiokande, which is on the order of $\mathcal{O}(10^{35})~\mathrm{years}$~\cite{Itow:2021rnc}. 
Note that this rough estimate of the proton lifetime could be modified if loop contributions are taken into account. 
This is because the $\bm{200}_F$ representation contains many heavy particles with masses at the unification scale, which contribute additively to the proton decay width.

\section{Experimental tests and its implication to fundamental theories}
Our model predicts the presence of sextet fermions at the energy scale of $\mathcal{O}(1-10)~\mathrm{TeV}$. 
Such exotic colored fermions can be pair produced through gluon fusion and quark-antiquark annihilations and 
searched for at the LHC~\cite{Carpenter:2021rkl} (see also~\cite{Han:2010rf, Richardson:2011df}). 
Since the interactions that induce their decays are suppressed by the unification scale, these sextet fermions are metastable. 
They form $R$ hadrons, which are bound states with quarks and gluons. At the LHC, these metastable particles can be detected via their large ionization energy loss, 
akin to the long-lived gluinos in supersymmetric models.
Although no significant deviations from the SM have been observed, upper bounds on the production cross section of long-lived gluinos have been derived~\cite{ATLAS:2016tbt, ATLAS:2016onr}. 
These bounds translate to a conservative mass limit of $1.58~\mathrm{TeV}$ for the sextet fermions in our model. 
Discrimination between sextet fermions and gluinos (octets) could be achieved through detailed calculations of their production cross sections.

Meanwhile, the quintuplet fermion dark matter can be probed via high-energy gamma-ray observations, such as Major Atmospheric Gamma-ray Imaging Telescope (MAGIC)~\cite{MAGIC:2022acl} 
and Cherenkov Telescope Array (CTA)~\cite{Montanari:2022buj}. 
The annihilation cross section of the dark matter is expected to be significantly enhanced due to Sommerfeld enhancement and bound-state formation effects~\cite{Mitridate:2017izz}. 
For instance, at the dark matter mass of $M_5=14~\mathrm{TeV}$ and a relative velocity of $v=2 \times 10^{-3}$ (assuming a Burkert density profile), 
the annihilation cross sections for $\chi^0\chi^0\to W^+W^-$ and $\gamma\gamma$ are predicted to be approximately $10^{-24}~\mathrm{cm^3/s}$ and $2\times 10^{-26}~\mathrm{cm^3/s}$, respectively. 
These values are within the detection capabilities of CTA~\cite{Montanari:2022buj}.

The quintuplet dark matter can also be probed through direct detection experiments. The strongest current constraint on the spin-independent cross section comes from LZ~\cite{tevpa2024lz}, 
with the bound of approximately $4.2\times 10^{-46}~\mathrm{cm^2}$ extrapolated to a dark matter mass of $14~\mathrm{TeV}$. 
This value is close to the theoretical prediction of $1.0\times 10^{-46}~\mathrm{cm^2}$~\cite{Farina:2013mla}.
Therefore, our model can be thoroughly tested via a combination of searches for exotic colored fermions and minimal dark matter in the $\mathcal{O}(10)~\mathrm{TeV}$ range.

These experimental searches could help explore fundamental theories of particle physics. 
It is well known that deriving a high dimensional representation, such as the quintuplet, from string theory is difficult~\cite{Baumgart:2024ezp}. 
So far, no explicit construction of an $SU(2)_L$ quintuplet from string theory has been found. 
Additionally, it is also known that if a high dimensional representation of gauge symmetries is included in string-based models, many lower dimensional representations typically accompany it. 
Based on this, all known string theories would be ruled out if future experiments confirm signals of minimal dark matter, 
necessitating modifications to existing string-based models to accommodate minimal dark matter.

\section{Additional comments}

In this work, we implicitly assumed that only the $SU(2)_L$ quintuplet in $\bm{200}_F$ and two pairs of $SU(3)_C$ sextets in the $\bm{15}_F$ and $\overline{\bm{15}}_F$ representations have masses of $\mathcal{O}(10)~\mathrm{TeV}$, while the other states are as heavy as the unification scale. In principle, achieving such large mass splittings is possible, though it requires fine-tuning. The Lagrangian relevant to the $\bm{200}_F$ fermion masses is given by 
\begin{align}
\mathcal{L}=&-\frac{M_{200}}{2}\bm{200}_F\bm{200}_F+\frac{Y_{200}}{2}\bm{24}_H\bm{200}_F\bm{200}_F.
\end{align}
After $SU(5)$ symmetry breaking via the vacuum expectation value of the singlet component in the adjoint Higgs $\bm{24}_H$, the masses of the components in $\bm{200}_F$ are split, 
with the mass splittings determined by the Clebsch-Gordan coefficients. 
The physical masses are then given by $M_i = M_{200} - C_{i} Y_{200} \langle \bm{24}_H\rangle$ where the index $i$ labels all the components in the $\bm{200}_F$ representation, as given in Eq.~(\ref{eq:200}). The Clebsch-Gordan coefficients $C_i$ are calculable using \texttt{GROUPMATH}~\cite{Fonseca:2020vke} and are listed in Table.~\ref{tab:1}.
Here, note that the coefficient for the quintuplet $(\bm{1},\bm{5})_{0}$ is normalized to $1$. 
Therefore, one can find that the required condition for only the $SU(2)_L$ quintuplet in the $\bm{200}_F$ representation to remain light is given by
$M_{200}-Y_{200}\langle \bm{24}_H\rangle = 14~\mathrm{TeV}$. The other states are as heavy as the unification scale. 
The same argument can also be applied to the pairs of $\bm{15}_F$ and $\overline{\bm{15}}_F$ fermions, with the Clebsch-Gordan coefficients listed in Table. 2.

\begin{table}[t]
\begin{center}
\begin{tabular}{c||cccc}\hline
Gauge symmetry & \multicolumn{3}{c}{Representation} & \\\hline
$SU(5)$ & $\bm{24}_H$ & $\bm{200}_F$ & $\bm{200}_F$ & $C_i$ \\\hline
\multirow{14}{*}{$G_{\mathrm{SM}}$}
        & $(\bm{1},\bm{1})_0$ & $(\bm{6},\bm{3})_{-5/3}$             & $(\overline{\bm{6}},\bm{3})_{5/3}$   & $1/6$\\
        & $(\bm{1},\bm{1})_0$ & $(\bm{15},\bm{2})_{-5/6}$            & $(\overline{\bm{15}},\bm{2})_{5/6}$  & $1/4$\\
        & $(\bm{1},\bm{1})_0$ & $(\bm{3},\bm{4})_{-5/6}$             & $(\overline{\bm{3}},\bm{4})_{5/6}$   & $-7/12$\\
        & $(\bm{1},\bm{1})_0$ & $(\bm{3},\bm{2})_{-5/6}$             & $(\overline{\bm{3}},\bm{2})_{5/6}$   & $-19/84$\\
        & $(\bm{1},\bm{1})_0$ & $(\bm{27},\bm{1})_{0}$               & $(\bm{27},\bm{1})_{0}$               & $2/3$\\
        & $(\bm{1},\bm{1})_0$ & $(\bm{8},\bm{3})_{0}$                & $(\bm{8},\bm{3})_{0}$                & $1/6$\\
        & $(\bm{1},\bm{1})_0$ & $(\bm{8},\bm{1})_{0}$                & $(\bm{8},\bm{1})_{0}$                & $1/14$\\
        & $(\bm{1},\bm{1})_0$ & $(\bm{1},\bm{5})_{0}$                & $(\bm{1},\bm{5})_{0}$                & $1$\\
        & $(\bm{1},\bm{1})_0$ & $(\bm{1},\bm{3})_{0}$                & $(\bm{1},\bm{3})_{0}$                & $11/21$\\
        & $(\bm{1},\bm{1})_0$ & $(\bm{1},\bm{1})_{0}$                & $(\bm{1},\bm{1})_{0}$                & $-2/7$\\
        & $(\bm{1},\bm{1})_0$ & $(\overline{\bm{3}},\bm{2})_{5/6}$   & $(\bm{3},\bm{2})_{-5/6}$             & $-19/84$\\
        & $(\bm{1},\bm{1})_0$ & $(\overline{\bm{3}},\bm{4})_{5/6}$   & $(\bm{3},\bm{4})_{-5/6}$             & $-7/12$\\
        & $(\bm{1},\bm{1})_0$ & $(\overline{\bm{15}},\bm{2})_{5/6}$  & $(\bm{15},\bm{2})_{-5/6}$            & $1/4$\\
        & $(\bm{1},\bm{1})_0$ & $(\overline{\bm{6}},\bm{3})_{5/3}$   & $(\bm{6},\bm{3})_{-5/3}$             & $1/6$\\\hline
\end{tabular}
\caption{Clebsch-Gordan coefficients for the components in $\bm{200}_F$ representation where the coefficient for $(\bm{1},\bm{5})_{0}$ is normalized to $1$.}
\label{tab:1}
\end{center}
\end{table}

\begin{table}[t]
\begin{center}
\begin{tabular}{c||cccc}\hline
$SU(5)$ & $\bm{24}_H$ & $\bm{15}_F$ & $\overline{\bm{15}}_F$ & $C_i$ \\\hline
\multirow{3}{*}{$G_\mathrm{SM}$}
 & $(\bm{1},\bm{1})_0$ & $(\bm{6},\bm{1})_{-2/3}$ & $(\bm{\overline{6}},\bm{1})_{2/3}$ & $1$\\
 & $(\bm{1},\bm{1})_0$ & $(\bm{3},\bm{2})_{1/6}$ & $(\overline{\bm{3}},\bm{2})_{-1/6}$ & $1/4$\\
 & $(\bm{1},\bm{1})_0$ & $(\bm{1},\bm{3})_{1}$ & $(\bm{1},\bm{3})_{-1}$ & $-3/2$\\\hline
\end{tabular}
\caption{Clebsch-Gordan coefficients for the components in pairs of $\bm{15}_F$ and $\overline{\bm{15}}_F$ where the coefficient for $(\bm{6},\bm{1})_{-2/3}$ is normalized to $1$.}
\label{tab:2}
\end{center}
\end{table}

Furthermore, the mass splitting problem could be resolved without fine-tuning via the Dimopoulos-Wilczek mechanism~\cite{Dimopoulos:1981xm, Srednicki:1982aj}, 
similar to the doublet-triplet splitting of the Higgs field. 
To achieve this, first, the $SU(5)$ gauge symmetry must be extended to $SO(10)$. 
Second, the $SU(2)_L$ quintuplet must be embedded in a complex representation of $SO(10)$ to forbid the mass term at tree level.
Namely, although the lowest $SO(10)$ representation containing the quintuplet is $\bm{660}$, it does not suffice. 
The lowest complex representation that includes the quintuplet is $\bm{2640}$. 
If a vacuum expectation value that breaks $SO(10)$ does not couple to the quintuplet, it can naturally remain light, 
though constructing a concrete model to realize this scenario would be challenging.

We also must consider the possible new interactions between the SM particles and the exotic sextet fermions. 
In particular, one can write the following gauge-invariant Yukawa couplings:
\begin{align}
\mathcal{L}\supset 
Y\bm{24}_H \bm{10}_F\bm{15}_F
+\tilde{Y}\bm{5}_H \bm{5}_F\bm{15}_F,
\end{align}
where the generation indices are omitted. 
We have checked that all additional couplings between the sextet fermions and quarks are suppressed by the unification scale.

It is well known that the Higgs potential in the SM becomes unstable at high energy scales. This issue persists in our model, as we have introduced only fermions and not scalars. 
However, this could be mitigated by introducing an additional gauge singlet scalar and mixed quartic couplings without affecting the successful gauge coupling unification 
discussed above~\cite{Cho:2021yue}.
Suppose the Higgs potential below the unification scale is given by 
\begin{align}
\mathcal{V}\supset \frac{\lambda_{HS}}{2} S^2|H|^2 + \frac{\lambda_S}{4!}S^4 + \frac{\lambda}{2}|H|^4,
\end{align}
where $S$ is a gauge singlet real scalar, which is $\bm{1}_H$ in the $SU(5)$ convention. 
The $\beta$ function for the quartic coupling $\lambda$ is modified by $\Delta b_{\lambda}=\lambda_{HS}^2$ at the one-loop level due to the presence of the singlet scalar. 
This positive contribution to the $\beta$ function stabilizes the Higgs potential at high energy scales, provided that the coupling strength is not too small. 

Even if we assume that the triplet $(\bm{1},\bm{3})_0$ in $\bm{200}_F$ is light instead of the quintuplet, it can also serve as stable dark matter, 
similar to the wino dark matter in supersymmetric models. 
In this case, gauge coupling unification can be achieved by adding a $(\bm{8},\bm{1})_0$ fermion and a $(\bm{1},\bm{3})_0$ scalar 
at the $\mathcal{O}(1)~\mathrm{TeV}$ scale~\cite{Ma:2005he, Cox:2016epl, Ma:2018uss}. 
These extra fields originate from the existing representations of $\bm{200}_F$ and the $\bm{24}_H$ adjoint Higgs in our model, respectively. 
The triplet fermion dark matter mass is predicted to be $M_3=2.7~\mathrm{TeV}$ in order to reproduce the observed relic abundance 
via the thermal freeze-out mechanism~\cite{Hisano:2006nn, Mitridate:2017izz}.
The octet fermion shares the same quantum numbers as the gluino in supersymmetric models, so we can apply the gluino mass bound to the triplet model. 
The current lower bound on the gluino mass is $2.3~\mathrm{TeV}$~\cite{ATLAS:2020syg, ATLAS:2021twp, ATLAS:2022zwa}.
For the $(\bm{1},\bm{3})_0$ fermion, the lower mass bound is obtained at the LHC and is $790~\mathrm{GeV}$\cite{ATLAS:2020wop}. 
For the triplet scalar, the relevant mass bound can be inferred from the charged Higgs boson decaying into $\Delta^{\pm} \to t\overline{b}$. 
This bound is approximately $1~\mathrm{TeV}$\cite{ATLAS:2021upq}. 
The charged scalar in the triplet can mix with the Goldstone boson absorbed by the $W$ boson in the SM via 
the cubic coupling $\bm{24}_H\bm{5}_H\overline{\bm{5}}_H$ after electroweak symmetry breaking.

\section{Summary}
The $SU(2)_L$ quintuplet fermion is the minimal dark matter candidate, and is embedded in a $\bm{200}$ representation of the nonsupersymmetric $SU(5)$ model, 
with a mass fixed at $14~\mathrm{TeV}$ to reproduce the thermal relic abundance. 
Gauge coupling unification is achieved without introducing new intermediate scales by adding two pairs of colored sextet fermions, 
which are embedded in the $\bm{15}$ and $\overline{\bm{15}}$ representations. 
It was found that the unification scale is very close to the reduced Planck scale, ensuring the stability of protons.
This model can be explored through comprehensive searches for the metastable exotic colored sextet fermions and $SU(2)_L$ quintuplet fermion dark matter at the $\mathcal{O}(1-10)~\mathrm{TeV}$ scale.
The colored sextet fermions can be probed at collider experiments if their mass is less than a few TeV. 
In principle, they can be distinguished from other exotic colored particles, such as gluinos in supersymmetric models or leptoquarks, 
due to differences in the magnitudes of their production cross sections. 
Additionally, the minimal dark matter can be searched for and experimentally identified by combining results from indirect and direct detection experiments, 
such as those from CTA and LZ, at the $14~\mathrm{TeV}$ scale.
If any signals of the quintuplet minimal dark matter are experimentally confirmed, it would indicate that known string theories must be appropriately modified, 
as there is no concrete example of deriving such a high dimensional representation from string theory in isolation.

\begin{acknowledgments}
This work was supported by a JSPS Grant-in-Aid for Scientific Research KAKENHI Grant No. 23H04004.
\end{acknowledgments}

\bibliographystyle{utphys}
\bibliography{reference}

\end{document}